\documentstyle[twocolumn,floats,aps,epsfig]{revtex}

\begin{document}
\bibliographystyle{try}


\def\isn{$^1$}
\def\sphn{$^2$}
\def\md{$^3$}
\def\jlab{$^4$}
\def\mitcamb{$^5$}
\def\mit{$^6$}
\def\rutgers{$^7$}
\def\bale{$^8$}
\def\lns{$^{9}$}
\def\ncat{$^{10}$}
\def\fiu{$^{11}$}
\def\ipn{$^{12}$}
\def\yerevan{$^{13}$}

\wideabs{
\title{Measurement of Tensor Polarization in Elastic Electron-Deuteron
Scattering at Large Momentum Transfer}

\author{D.~Abbott,\jlab\   
	A.~Ahmidouch,\mit$^,$\ncat\  
	H.~Anklin,\fiu\   
	J.~Arvieux,\lns$^,$\ipn\   
	J.~Ball,\sphn$^,$\lns\   
	S.~Beedoe,\ncat\   
	E.J.~Beise,\md\   
	L.~Bimbot,\ipn\    
	W.~Boeglin,\fiu\   
	H.~Breuer,\md\   
	P.~Brindza,\jlab\ 
	R.~Carlini,\jlab\    
	N.S.~Chant,\md\   
	S.~Danagoulian,\ncat$^,$\jlab\  
	K.~Dow,\mit\   
	J.-E.~Ducret,\sphn\    
	J.~Dunne,\jlab\
	L.~Ewell,\md\   
	L.~Eyraud,\isn\   
	C.~Furget,\isn\   
	M.~Gar\c con,\sphn\    
	R.~Gilman,\jlab$^,$\rutgers\   
	C.~Glashausser,\rutgers\   
	P.~Gueye,\jlab\   
	K.~Gustafsson,\md\   
	K.~Hafidi,\sphn\    
	A.~Honegger,\bale\    
	J.~Jourdan,\bale\   
	S.~Kox,\isn\   
	G.~Kumbartzki,\rutgers\   
	L.~Lu,\isn\    
	A.~Lung,\md\   
	D.~Mack,\jlab\   
	P.~Markowitz,\fiu\   
	J.~McIntyre,\rutgers\   
	D.~Meekins,\jlab\   
	F.~Merchez,\isn\    
	J.~Mitchell,\jlab\
	R.~Mohring,\md\    
	S.~Mtingwa,\ncat\   
	H.~Mrktchyan,\yerevan\   
	D.~Pitz,\sphn$^,$\md$^,$\jlab\    
	L.~Qin,\jlab\    
	R.D.~Ransome,\rutgers\   
	J.-S.~R\'eal,\isn\   
	P.G.~Roos,\md\    
	P.~Rutt,\rutgers\    
	R.~Sawafta,\ncat$^,$\jlab\ 
	S.~Stepanyan,\yerevan\      
	R.~Tieulent,\isn\   
	E.~Tomasi-Gustafsson,\sphn$^,$\lns\    
	W.~Turchinetz,\mitcamb\   
	K.~Vansyoc,\jlab\    
	J.~Volmer,\jlab\   
	E.~Voutier,\isn\   
	W.~Vulcan,\jlab\    
	C.~Williamson,\mitcamb\   	
	S.A.~Wood,\jlab\    
	C.~Yan,\jlab\   
	J.~Zhao,\bale\   
	and 
	W.~Zhao\mitcamb\\
        (The Jefferson Lab t$_{\hbox{20}}$ collaboration)
       }

\address{
\isn Institut des Sciences Nucl\'eaires, IN2P3-UJF, 38026 Grenoble, France\\
\sphn DAPNIA/SPhN, CEA/Saclay, 91191 Gif-sur-Yvette, France\\ 
\md University of Maryland, College Park, MD 20742, USA\\
\jlab Thomas Jefferson National Accelerator Facility, Newport News, VA 23606, USA\\
\mitcamb M.I.T.-Laboratory for Nuclear Science and Department of Physics, Cambridge, MA 02139, USA\\
\mit M.I.T.-Bates Linear Accelerator, Middleton, MA 01949, USA\\
\rutgers Rutgers University, Piscataway, NJ 08855, USA\\
\bale Basel Institut f\"{u}r Physik, Switzerland\\
\lns LNS-Saclay,  91191 Gif-sur-Yvette, France\\
\ncat North Carolina A. \& T. State University, Greensboro, NC 27411, USA\\
\fiu Florida International University, Miami, FL 33199, USA\\
\ipn IPNO, IN2P3, BP 1, 91406 Orsay, France\\
\yerevan Yerevan Physics Institute, 375036 Yerevan, Armenia\\
}

\maketitle

\begin{abstract}
Tensor polarization observables ($\mbox{t}_{20}$, $\mbox{t}_{21}$ and
$\mbox{t}_{22}$) have been
measured in elastic electron-deuteron scattering for six values of 
momentum transfer between 0.66 and 1.7~(GeV/c)$^2$. The experiment was performed
at the Jefferson Laboratory in Hall C using the electron HMS Spectrometer,
a specially designed deuteron magnetic channel and 
the recoil deuteron polarimeter POLDER.
The new data determine to much larger $Q^2$ the deuteron charge form
factors $G_C$ and $G_Q$. They are in good agreement with relativistic
calculations and disagree with pQCD predictions.
\end{abstract}



} 

\narrowtext

The development of a quantitative understanding of the structure
of the deuteron, the only two-nucleon bound state, has long been
considered an important testing ground for models of the 
nucleon-nucleon potential. Nevertheless, the charge distribution 
of the deuteron is not well known experimentally, because it is only 
through the use of both polarization measurements and unpolarized
elastic scattering cross sections that it
can be unambiguously determined. In the experiment described here,
a precise determination of the charge form factor of the deuteron
is presented
through measurement of the deuteron tensor polarization observables
up to a momentum transfer of $Q^2$=1.7~(GeV/c)$^2$, for the first time
well beyond its zero crossing.

Since the deuteron is a spin-1 nucleus, its
electromagnetic structure is described by three form factors:
the charge monopole $G_C$, quadrupole $G_Q$ and magnetic dipole
$G_M$. Thus it is possible to unambiguously separate the three
components only through measurement of three observables. 
In the one-photon exchange approximation, the elastic scattering 
cross section is typically
expressed in terms of structure functions $A(Q^2)$ and $B(Q^2)$
($d\sigma/d\Omega\propto{\cal S}$ with ${\cal S}=A(Q^2)+B(Q^2)\tan^2({\theta_e/2})$,
see full expressions e.g. in \cite{aqhallc}) that can be
separately determined by variation of the scattered electron angle $\theta_e$ 
for a given momentum transfer $Q^2$ to the deuteron.

The third observable can be
the cross section dependence on deuteron (tensor or vector)
polarization. The tensor analyzing powers can 
be measured using a polarized deuteron target (with unpolarized beam)
\cite{Dmitriev85,Gilman90,Ferro-Luzzi96,Bouwhuis99}. 
Alternatively, 
the tensor moments of the outgoing deuterons can be measured
using unpolarized beam and target\cite{Schulze84,Garcon94}.  
Both types of experiment result in the same combinations of form factors:
\begin{eqnarray}
\nonumber
\mbox{t}_{20} &=& -\frac{1}{\sqrt{2}\ {\cal S}} 
\left(\ {8 \over 3} \eta G_C G_Q + {8 \over 9} \eta^2 G_Q^2 \right.  \\
&& +\left. \frac{1}{3} \eta \left[1+2(1+\eta)\tan^2 \frac{\theta_e}{2}  
\right]G_M^2 \right) \label{eq:obst20}\\
\mbox{t}_{21} &=&  \frac{2}{\sqrt{3}\ {\cal S} \cos \frac{\theta_e}{2}}\ \eta 
\left[\eta + \eta^2 \sin^2 \frac{\theta_e}{2} \right]^{\frac{1}{2}} 
G_M G_Q \label{eq:obst21}\\
\mbox{t}_{22} &=&  -{1 \over 2\sqrt{3}\  {\cal S}}\ \eta G_M^2\;,\ \ \ \ 
\mbox{with $\eta=Q^2/4M_d^2$.} 
\label{eq:obst22}
\end{eqnarray}

The tensor moment $\mbox{t}_{20}$\ is particularly interesting due to its
sensitivity to $G_C$.
It has been previously measured using either the polarimeter or polarized
target technique, up to 0.85~(GeV/c)$^2$.
In our experiment described 
below, new measurements of t$_{2q}$ were performed between 0.66 and 1.7~(GeV/c)$^2$.
$A(Q^2)$ was measured previously up to 4~(GeV/c)$^2$,
but with significant discrepancy between data sets in 
our $Q^2$ range\cite{Cramer85,Elias69,Arnold75}.
New $A(Q^2)$ data \cite{aqhallc,aqhalla}, including some from this experiment,
resolve many of these discrepancies.
$B(Q^2)$, which is typically a factor 10 smaller than $A(Q^2)$, 
has been measured up to 2.8~(GeV/c)$^2$\cite{Bosted90}.

Our experiment was performed at the Thomas Jefferson National Accelerator Facility 
(JLab) in
the experimental Hall C.
A continuous electron beam with a typical current
between 80 and 120~$\mu$A was used together with a 12~cm long
liquid deuterium target resulting in an average luminosity of about
3$\times$10$^{38}$~cm$^{-2}$s$^{-1}$.

The scattered electrons were detected in the
High Momentum Spectrometer (HMS), in coincidence with the recoil deuterons.
The scattered deuterons were transported by a specially designed magnetic channel
composed of warm magnets, three quadrupoles and one dipole,
to the POLDER
polarimeter. This magnetic channel
optimized the acceptance matching between 
the two arms, which varied from 0.5 to 1 depending on the kinematics,
and focussed the elastically scattered deuterons on 
the target of POLDER.
The deuteron magnetic channel was set at 
a fixed angle of 60.5$^{\circ}$. The six different $Q^2$ values were then
obtained by changing both
the beam energy (from 1.4 to 4~GeV) and the detection angle of the HMS 
spectrometer.

The elastic scattering events were selected by setting cuts on
the primary vertex position and $\gamma^*-d$ invariant mass, as determined by the HMS,
the particle energy loss in two thin plastic scintillators located before the polarimeter target, 
and the time coincidence measurement between the two arms.
The combination of these redundant selection criteria reduced
the contribution of remaining background (mainly coming from
random coincidences between electrons and protons, and from coherent pion production)
to less than 0.2\%.

\begin{figure}
  \begin{center}
  \epsfig{file=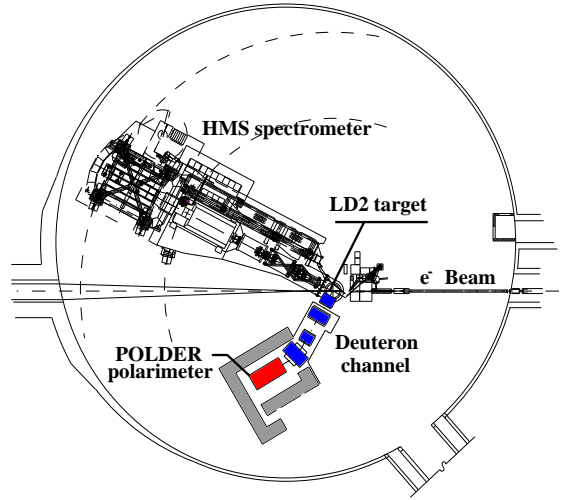,width=0.85\columnwidth}
  \end{center}
  \caption{Experimental set-up of the $\mbox{t}_{20}$ experiment in Hall~C at TJNAF.}
  \label{fig:setup}
\end{figure}

\begin{table*}
   \caption{Measured tensor Polarization observables t$_{kq}(\theta_e)$, 
    with statistical and systematic errors.
    The charge form factors are given, with in some occurences asymmetric overall
   errors.}
   \label{tab:t20res}
	\begin{tabular}{ccccccc}
	$Q^2$ (GeV/c)$^2$  &  .651   &  .775   & 1.009   & 1.165   & 1.473   & 1.717   \\ \hline
	$\theta_e$ (deg.)  & 35.6   & 33.4   & 29.8   & 27.3   & 23.0   & 19.8   \\
	\def\obs{\mbox{t}_{20}}
	$\obs$\begin{tabular}{l}\scriptsize$\pm\Delta stat.$\\\scriptsize$\pm\Delta syst.$\end{tabular}&
	-.546 \begin{tabular}{l}\scriptsize$\pm$.038\\\scriptsize$\pm$.170\end{tabular}&
	-.322 \begin{tabular}{l}\scriptsize$\pm$.031\\\scriptsize$\pm$.088\end{tabular}&
	 .191 \begin{tabular}{l}\scriptsize$\pm$.034\\\scriptsize$\pm$.043\end{tabular}&
	 .301 \begin{tabular}{l}\scriptsize$\pm$.048\\\scriptsize$\pm$.056\end{tabular}&
	 .625 \begin{tabular}{l}\scriptsize$\pm$.094\\\scriptsize$\pm$.141\end{tabular}&
	 .477 \begin{tabular}{l}\scriptsize$\pm$.178\\\scriptsize$\pm$.063\end{tabular} \\
	\def\obs{\mbox{t}_{21}}
	$\obs$\begin{tabular}{l}\scriptsize$\pm\Delta stat.$\\\scriptsize$\pm\Delta syst.$\end{tabular}&
	 .463 \begin{tabular}{l}\scriptsize$\pm$.051\\\scriptsize$\pm$.113\end{tabular}&
	 .315 \begin{tabular}{l}\scriptsize$\pm$.041\\\scriptsize$\pm$.083\end{tabular}&
	 .201 \begin{tabular}{l}\scriptsize$\pm$.042\\\scriptsize$\pm$.077\end{tabular}&
	 .220 \begin{tabular}{l}\scriptsize$\pm$.056\\\scriptsize$\pm$.094\end{tabular}&
	 .166 \begin{tabular}{l}\scriptsize$\pm$.096\\\scriptsize$\pm$.056\end{tabular}&
	-.001 \begin{tabular}{l}\scriptsize$\pm$.152\\\scriptsize$\pm$.058\end{tabular} \\
	\def\obs{\mbox{t}_{22}}
	$\obs$\begin{tabular}{l}\scriptsize$\pm\Delta stat.$\\\scriptsize$\pm\Delta syst.$\end{tabular}&
	 .087 \begin{tabular}{l}\scriptsize$\pm$.042\\\scriptsize$\pm$.037\end{tabular}&
	-.027 \begin{tabular}{l}\scriptsize$\pm$.030\\\scriptsize$\pm$.037\end{tabular}&
	-.018 \begin{tabular}{l}\scriptsize$\pm$.029\\\scriptsize$\pm$.029\end{tabular}&
	 .022 \begin{tabular}{l}\scriptsize$\pm$.035\\\scriptsize$\pm$.037\end{tabular}&
	-.023 \begin{tabular}{l}\scriptsize$\pm$.054\\\scriptsize$\pm$.048\end{tabular}&
	-.133 \begin{tabular}{l}\scriptsize$\pm$.074\\\scriptsize$\pm$.047\end{tabular} \\ \hline
	$G_C \times 10^{2}$&
	-.117 $\pm$ .162&
	-.253 $\pm$ .063&
	-.396 $\pm$ .028&
	-.348 $\pm$ .031&
	-.310 \begin{tabular}{l}\scriptsize $+$.053\\\scriptsize $-$.061\end{tabular}&
	-.194 \begin{tabular}{l}\scriptsize $+$.036\\\scriptsize $-$.052\end{tabular}\\
	$G_Q$&
	 .393 $\pm$ .010&
	 .259 $\pm$ .007&
	 .122 $\pm$ .004&
	 .080 $\pm$ .003&
	 .034 \begin{tabular}{l}\scriptsize $+$.005\\\scriptsize $-$.007\end{tabular}&
	 .023 \begin{tabular}{l}\scriptsize $+$.002\\\scriptsize $-$.004\end{tabular}\\
	\end{tabular}
\end{table*}

The polarimeter POLDER\cite{Polder94,Eyraud98} is based on the charge
exchange reaction
$^1$H($\vec{d}\,$,2p)n, which provides
sizeable angular asymmetries depending on the tensor, but not on the vector,
components of the incident deuteron polarization \cite{Carbonell91}.
The direction of deuterons is measured with two multi-wire proportional
chambers placed upstream of a 22~cm long
liquid hydrogen target.
Deuterons that undergo a charge exchange reaction, produce
two outgoing protons with small relative angle and momentum
in the forward
direction. They are detected, and their positions measured, in
two hodoscopes, composed of plastic scintillator bars. The polar and azimuthal angle
distributions of the center of mass of the two protons are used to determine
the deuteron beam polarization.
The efficiency of the polarimeter, defined as the fraction of the deuterons
undergoing a charge exchange reaction, is of the order of (3--6)$\times$10$^{-3}$
and must be measured with a precision of 1\%.
The absolute polarized efficiency $\epsilon _{pol}(\theta,\varphi)$ of the polarimeter,
measured in 
this experiment, has to be compared to the unpolarized value 
$\epsilon _{0}(\theta)$ through the relation~:
\begin{eqnarray}
  \nonumber
   && \epsilon_{pol} \left( \theta,\varphi \right) =
  \epsilon _{0} (\theta) \left[ \frac{}{} 1 + \mbox{t}_{20}\; \mbox{T}_{20} (\theta) \right.\\
  &+& \left. 2\cos(\phi)\;\mbox{t}_{21}\;\mbox{T}_{21}(\theta)+2\cos(2\phi)\;\mbox{t}_{22}\;
  \mbox{T}_{22}(\theta)\frac{}{}\right]\;,
     \label{eq:efficiency}
\end{eqnarray}

\noindent where T$_{kq}$ are the analyzing powers of the $^1$H($\vec{d}\,$,2p)n reaction,
t$_{kq}$ the deuteron polarization coefficients to be determined in this experiment,
$\theta$ is the angle
between the incident deuteron and the proton pair momentum and $\phi$ the angle 
between the normal to the $^1$H($\vec{d}\,$,2p)n reaction plane and the
$e$-$d$ scattering plane. 

The analyzing powers and the unpolarized efficiency 
were measured previously at SATURNE using deuteron beams of known polarization 
in the range of kinetic energies between 140 and 520~MeV, in 10 to 30~MeV steps \cite{Eyraud98}.
The polarimeter data analysis was identical
for the calibration and the JLab measurements.
The selection of charge exchange events was achieved by requiring
a coincidence between the detection of
one incident particle before the target
and the detection of two charged particles in the hodoscopes.
Events with several incident particles were rejected using cuts
on the energy loss measured in the scintillators and on the multiplicity information
from the wire chambers.
Time of flight was measured between the incident deuteron
and the hit bars of hodoscopes. Cuts on this time of flight,
together with an algorithm to reconstruct proper proton
tracks, led to a clean selection of charge exchange events.
Two different tracking algorithms, with different geometrical selection
criteria, were used to prove that the background 
(parasitic reactions in the polarimeter or multiple
incident particles not rejected by the front end of the polarimeter)
within the charge exchange events was negligible.
The angles $\theta$ and $\varphi$ were then calculated using the direction of
the deuteron and the proton tracks.
The deduced efficiency was then stable  within 0.6\% under changes of 
experimental conditions (except for 
the data at the lowest deuteron energy where variations reached 1.2\%).

The distributions of incident deuteron energy on the polarimeter at JLab had
a large width,
16 to 51~MeV, and were not centered at any of the energies of the calibration experiment.
The observables $\epsilon _{0}$ and T$_{kq}$ of Eq.~\ref{eq:efficiency}
were then obtained by weighting with deuteron energies the interpolated
SATURNE data. For this procedure, the deuteron energy was calculated for each
event from the JLab beam
energy and the scattered electron angle, with a correction coming from energy
loss (mostly in the LD$_2$ target).

The tensor polarization observables were obtained from
Eq.~\ref{eq:efficiency} through a minimization procedure, adjusting the
$\mbox{t}_{2q}$ values
such that the angular distribution
on the right-hand side best reproduced the angular distribution of
the polarized efficiency measured in this
experiment.
In this fit, the resulting value of $\mbox{t}_{20}$ is highly correlated with
the fixed value of $\epsilon _{0}$, but is uncorrelated with $\mbox{t}_{21}$
and $\mbox{t}_{22}$.
A small spin precession correction was then applied, corresponding
to a net deviation of 29.7$^{\circ}$ in the deuteron channel.
Our results \cite{Eyraud98,Gustafsson00,Hafidi99,Zhao99} are given in Table~I.
The systematic errors include those due to analysis cuts
(mostly from geometrical POLDER cuts), the uncertainties in the deuteron energy
(from beam energy, electron angle, beam position on target),
the uncertainties in calibration results (statistical and systematic errors on analyzing powers,
interpolation, absolute stability on unpolarized efficiency) as well as the small instrumental
unphysical asymmetries measured in the calibration.
The uncertainty coming from the knowledge of the deuteron energy 
as well as the one due to calibration results
were larger at the lowest
$Q^2$ points because of the energy dependence of $\epsilon _{0}$ and the stability
of the polarimeter at this deuteron energies.
In the case of the point at 1.47~(GeV/c)$^2$, the $\theta$ distribution
of $\epsilon_{pol}$ did not match exactly the expected behaviour from
Eq.~\ref{eq:efficiency}. This led to the addition of a contribution
to the systematic error in t$_{20}$ for this point of $\Delta t_{20}$=0.1.
These systematic errors were combined quadratically and are mostly uncorrelated
for the different data points.

\begin{figure}[t]
   \epsfig{figure=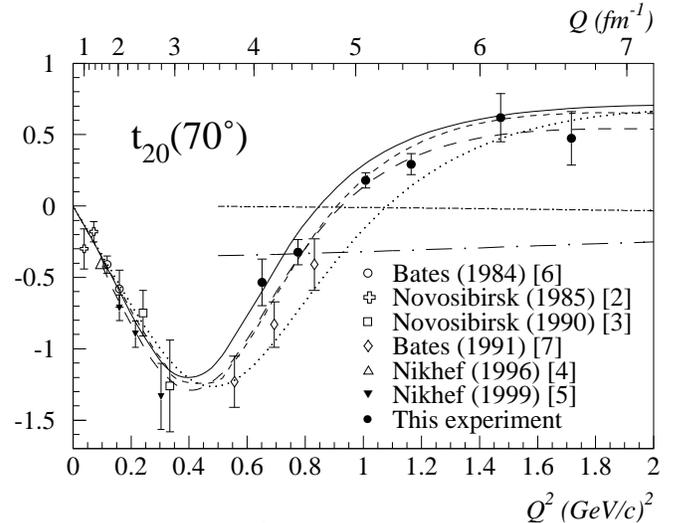,width=\columnwidth}
   \caption{\small t$_{20}$ at $\theta_e$=70$^{\circ}$
   compared to theoretical predictions; dotted line (NRIA) and full line
   (NRIA+MEC+RC)\protect \cite{Wiringa95};
   relativistic models with dashed line\protect \cite{Phillips99} and
   long dashed line\protect \cite{Carbonell99}; 
   pQCD calculations with dashed-dotted line\protect \cite{Brodsky92}
   and long dashed-dotted line\protect \cite{Kobushkin94}.}
   \label{fig:t20res}
\end{figure}

\begin{figure}[t]
   \epsfig{figure=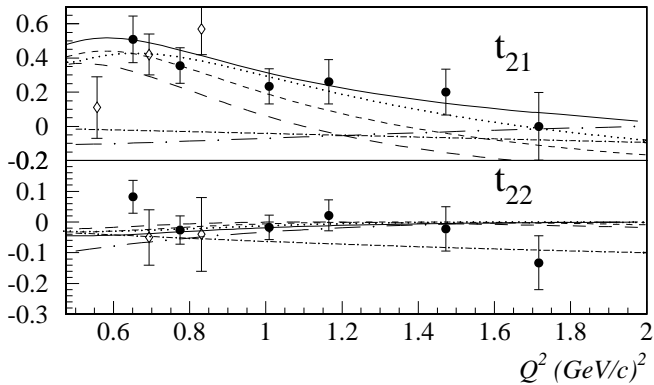,width=\columnwidth}
   \caption{t$_{21}$ and t$_{22}$ at $\theta_e$=70$^{\circ}$.
	See Fig.~\ref{fig:t20res} and text for the curves.}
   \label{fig:tijres}
\end{figure}

For the sake of comparison with other data and with theoretical models,
small corrections (of order $B/A$ and $B\tan^2(\theta_e/2)/A$,
see Eqs.~\ref{eq:obst20}--\ref{eq:obst22}) were applied
to calculate t$_{2q}$ at the conventionally accepted angle of 70$^{\circ}$. 
These results obtained for the tensor polarization observables
are shown in Fig.~\ref{fig:t20res} and  \ref{fig:tijres},
and compared with the existing world data
\cite{Dmitriev85,Gilman90,Ferro-Luzzi96,Bouwhuis99,Schulze84,Garcon94}
and with several recent theoretical predictions.
The error bars include both 
statistical and systematic errors, combined quadratically.

Where the new data overlap with the earlier Bates data\cite{Garcon94},
they agree within the combined uncertainties, although it 
appears that the Bates $\mbox{t}_{20}$ values are
systematically more negative.
The indication of t$_{21}$ crossing 0 is consistent with the existence of
a node of the magnetic form factor $G_M$ (see
Eq. \ref{eq:obst21}) around 2~(GeV/c)$^2$, as first indicated by a
measurement of $B(Q^2)$ \cite{Bosted90}.

\begin{figure}[t]
   \epsfig{figure=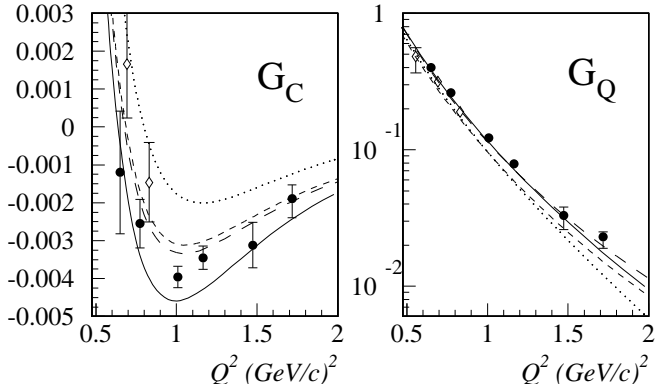,width=\columnwidth}
   \caption{Monopole (G$_C$) and quadrupole (G$_Q$) charge form factors of
	the deuteron.
	See Fig.~\ref{fig:t20res} and text for the curves and references.}
   \label{fig:gcgqres}
\end{figure}

A recent non-relativistic impulse approximation prediction
(NRIA)\cite{Wiringa95}
calculated using the Argonne $v_{18}$ potential for the {\em NN} interaction,
seems to reproduce the Bates data
(the dotted curve in Fig.\ref{fig:t20res} and \ref{fig:tijres}).
But to be in a reasonable agreement with our
new t$_{20}$ data, meson exchange currents (MEC) and relativistic
corrections (RC) (solid curve) must be included.
The MEC calculation includes pair
terms and the $\rho\pi\gamma$ mechanism,
for which the strength is not well known\cite{VanOrden95}.

Two relativistic and covariant models, both including MEC,
are compared with the data.
The dashed curve\cite{Phillips99} uses a
three-dimensional reduction of the Bethe-Salpeter equation
using an equal-time formalism, and includes 
$\rho\pi\gamma$ exchange currents. 
The long dashed curve\cite{Carbonell99} is the prediction of a model
developed in the framework of 
the explicitly covariant version of light front dynamics.
It uses a full relativistic potential, calculated
with the same set of mesons and parameter values used in the construction of the
Bonn potential, but does not include the $\rho\pi\gamma$ MEC.
Both models are in good agreement with
our t$_{20}$ data, but the prediction based
on the light cone formalism agrees better with the last NIKHEF data,
at lower $Q^2$\cite{Bouwhuis99}.
However, this model does not
reproduce the position of the node of $G_M$, which leads to a bad 
description of t$_{21}$.

Finally, two pQCD calculations, predicting simple
relations between the form factors of the deuteron, are shown by
dashed-dotted curves in Fig.~\ref{fig:t20res} and \ref{fig:tijres}.
One of them\cite{Brodsky92} uses only the helicity-conserving
matrix element of the electromagnetic current, arguing that it should dominate
above 1~(GeV/c)$^2$. The other one\cite{Kobushkin94} includes
the helicity-one-flip matrix element and fixes its contribution using
the location of the node of $B(Q^2$), taken to be at 2~(GeV/c)$^2$.
Comparison with t$_{20}$ and t$_{21}$ measurements clearly
shows that both pQCD predictions fail to reproduce our data
contrary to the scale in four-momentum transfer given by the authors for the
applicability of their calculations. 

Deuteron form factors can be expressed in terms of $A$
(which have been interpolated using the latest data \cite{aqhallc,aqhalla}
in our  $Q^2$ range), $B$ and $\mbox{t}_{20}$.
These equations are quadratic and admit,
in general, two solutions.
Ambiguities in the choice of the proper solution remain only for our
two highest $Q^2$ points,
due to the fact that $\mbox{t}_{20}$ is close to its maximum,
where the two solutions are nearly degenerate. If we follow the prediction of
most theoretical
models, according to which the maximum of $\mbox{t}_{20}$ occurs beyond 
our highest $Q^2$ point,
one of the two solutions can be selected. This particular issue will be
addressed elsewhere
in more detail~\cite{phen_ff}. The errors in $G_C$ (see Table~I)
come predominantly
from the errors in the $\mbox{t}_{20}$ measurements.

The results for the charge form factors $G_C$ and $G_Q$, shown in
Fig.~\ref{fig:gcgqres}, lead to the same conclusions
made for $\mbox{t}_{20}$ data about the models and the Bates data.
The results for the charge form factor $G_C$
show a node located at a
lower value than inferred from the previous Bates data.
This removes the inconsistency, pointed out by Henning \cite{henning95},
in the location of the
minimum for the charge form factor of two- and three-nucleon systems.
Our data also suggest for the first time a secondary maximum of $|G_C|$.
The height
of this maximum seems to be inconsistent with that of the corresponding
three-nucleon system, within the same non-relativistic models~\cite{henning95}.

In summary, we have measured the tensor polarization observables
in electron deuteron elastic scattering between 0.65 and
1.72~(GeV/c)$^2$.
Our data on t$_{20}$, used in conjunction with data on the structure 
function $A(Q^2$), provide a determination of the charge and
quadrupole form factors.
We have compared our results with only few recent calculations.
Within non-relativistic models,
all the observables are in favor of the inclusion of meson exchange currents
and relativistic effects in the theoretical calculations. In fact the present
data could constitute the best experimental determination of isoscalar meson 
exchange currents. Recent relativistic models are in remarkable agreement
with our data.
Finally, the $Q^2$ range covered by these data
shows that the pQCD predictions are not reliable for these
momentum transfers.

{\it Acknowledgements:} 
We acknowledge the outstanding work of the JLab accelerator division
and the Hall C engineering staff. 
We thank the Indiana University
Cyclotron Facility for its technical help.
This work was supported by 
the French Centre National de la Recherche Scientifique and
Commissariat \`a l'Energie Atomique,
the U.S. Department of Energy
and National Science Foundation,
the Swiss National Science Foundation,
and the K.C. Wong Foundation.

\end{document}